\documentclass[sigconf]{acmart}

\usepackage{booktabs} 
\usepackage[english]{babel} 
\usepackage[utf8x]{inputenc} 
\usepackage[colorinlistoftodos]{todonotes} 

\setcopyright{rightsretained}

\begin{document}
\title{LivingLab PJAIT: Towards Better Urban Participation of Seniors}


\author{Wies\l{}aw Kope\'{c}}
\affiliation{%
  \institution{Polish-Japanese Academy of Information Technology}
  \streetaddress{86 Koszykowa str.}
  \postcode{02-008}
  \city{Warsaw}
  \country{Poland}
}
\email{wieslaw.kopec@pja.edu.pl}

\author{Kinga Skorupska}
\affiliation{%
  \institution{Polish-Japanese Academy of Information Technology}
  \streetaddress{86 Koszykowa str.}
  \postcode{02-008}
  \city{Warsaw}
  \country{Poland}
}
\email{kinga.skorupska@pja.edu.pl}

\author{Anna Jaskulska}
\affiliation{%
  \institution{Polish-Japanese Academy of Information Technology}
  \streetaddress{86 Koszykowa str.}
  \postcode{02-008}
  \city{Warsaw}
  \country{Poland}
}
\email{anna.jaskulska@pja.edu.pl}

\author{Katarzyna Abramczuk}
\affiliation{%
  \institution{Warsaw University}
  \streetaddress{Krakowskie Przedmie\'{s}cie 26/28}
  \postcode{00-927}
  \city{Warsaw}
  \country{Poland}
}
\email{k.abramczuk@uw.edu.pl}

\author{Radoslaw Nielek}
\affiliation{%
  \institution{Polish-Japanese Academy of Information Technology}
  \streetaddress{86 Koszykowa str.}
  \postcode{02-008}
  \city{Warsaw}
  \country{Poland}
}
\email{nielek@pja.edu.pl}

\author{Adam Wierzbicki}
\affiliation{%
  \institution{Polish-Japanese Academy of Information Technology}
  \streetaddress{86 Koszykowa str.}
  \postcode{02-008}
  \city{Warsaw}
  \country{Poland}
}
\email{adamw@pja.edu.pl}

\renewcommand{\shortauthors}{W. Kopec et al.}

\begin{abstract}
In this paper we provide a brief summary of development LivingLab PJAIT as an attempt to establish a comprehensive and sustainable ICT-based solution for empowerment of elderly communities towards better urban participation of seniors. We report on our various endeavors for better involvement and participation of older adults in urban life by lowering ICT barriers, encouraging social inclusion, intergenerational interaction, physical activity and engaging older adults in the process of development of ICT solutions. We report on a model and assumptions of the LivingLab PJAIT as well as a number of activities created and implemented for LivingLab participants: from ICT courses, both traditional and e-learning, through on-line crowdsourcing tasks, to blended activities of different forms and complexity. We also provide conclusions on the lessons learned in the process and some future plans, including solutions for better senior urban participation and citizen science.

\end{abstract}


\copyrightyear{2017} 
\acmYear{2017} 
\setcopyright{acmlicensed}
\acmConference{WI '17}{August 23-26, 2017}{Leipzig, Germany}\acmPrice{15.00}\acmDOI{10.1145/3106426.3109040}
\acmISBN{978-1-4503-4951-2/17/08}





\maketitle



\section{Introduction}

As determined by Eurostat, in 2014 older people, aged 65+ constituted 18.5 \% of the entire EU-28 population, which is a 4\% increase when compared to the 1994 statistics. This is in line with trends observed in most developed countries, with Japan being the outlier as the most aged society by 2050\footnote{$http://www.un.org/en/development/desa/population/publications/pdf/ageing/WPA2015\\_Report.pdf$} . The United Nations estimate that in 2050 this figure will reach a global average of 20 \% and with the increasing lifespan and medical advancements the number of people over 80 will double.\cite{UNreport} What follows is that the current old-age dependency ratio, which for EU-28 stands at almost 29 \%, meaning that there are about four working adults for every senior, will further increase.

On one hand this presents a difficulty, as with the current tendency to exclude the elderly from some aspects of social participation, their voice is unheard and their needs are unmet, but on the other hand it is an opportunity to tap into their potential and resources, not only to tackle their own problems, but also to allow them to help solve the issues common to all groups in the society. 

This requires the development of an approach to allow the elderly to actively participate in the society - which increasingly relies on ICT skills - and to contribute their time and experience to solving the problems of aging societies, including a shortage of resources and overpopulation. This can be done by crowdsourcing insights from this age group that could be studied to develop better services and by involving them in the creation of the solutions themselves, which is facilitated by the Living Lab framework.

For this, seniors need to learn ICT skills and ways to collaborate in intergenerational groups. As we have observed, one of the most effective ways to achieve this is to offer a mix of online and offline activities and to teach ICT skills in reference to the real world, people and locations. Owing to the use of location based services and augmented-reality the interaction with mobile devices in urban settings can turn seniors into engaged users,  willing contributors, and eventually designers in the technological marketplace.

Because the design of public spaces is now of key importance, both due to the fast progressing urbanization and the high population density, at Living Lab at PJAIT, we propose to tap into the potential of senior crowdsourcing solutions in public space design. The elderly are the demographic that has the knowledge and resources to offer powerful insights into the changes needed to make public places serve their intended purpose as well as facilitate engagement and encourage physical activity outdoors. In this article we will report our activities within the Living Lab framework which, step by step, build our experience and potential towards greater social inclusion of older adults into the city life. This will be facilitated by gaining insights into the senior ICT-skill learning process, their motivation and by creating a senior community of learners, users, contributors, testers and designers.

The rest of the article is organized as follows. First we present related works, which is followed by a brief description of PJAIT LivingLab  framework project. Next, we present details on two fundamental parts of the LivingLab platform: e-courses and crowdsourcing tasks, followed by brief overview of different activites with participants of the PJAIT LivingLab . Finally, we discuss our findings and present plans for the future.

\section{Related Works}
As it was outlined above in the previous section facing the aging society problems should be strictly connected with the process of enhancing the participation of seniors in the urban life and their involvement in creation process of new solutions. There are many ways to achieve that goal, however, two major ones are closely related to the topic of the INTENSE special session of WI-IAT 2017 conference. 

The first topic is related to the interaction with smart urban objects and the second, to the development of devices and software connected to the crucial topics of safe and independent living in urban space. While the former is focused on current trends in smart cities, namely sensors and IoT, that are subject of investigation from many authors, including Koch \cite{kotteritzsch2016expand}, the latter is strictly connected to our approach of involving the older adults in the process of active participation in developing solutions which is connected with one of the fundamental concepts of contemporary design studies: participatory design.

\subsection{Human-centered approach}

The potential of participatory design, that is designing together with the end-users, is increasingly recognized as crucial in developing well-targeted solutions. This approach is also related in literature i.e. by Sanders \cite{sanders2008co} to another term known as \textit{co-design}. Although these terms have different origins, they refer to the same human-centered approach, which is widely used not only in software engineering, but also in fields ranging from architecture to the healthcare industry \cite{szebeko2010co}. These concepts put human beings at the center of the design process, although there is a small but significant distinction between traditional user-centered design and participatory design: the former refers to the process of designing for users, while the latter is related to design process with potential end-users \cite{sanders2002user,sanders2008co}. According to Sanders the next step in participatory design is connected with the process of designing by the users themselves. She also claims that participatory approach can be applied throughout the whole design cycle, even including the pre-design stage. On the other hand Ladner  \cite{ladner2015design} limits the definition of the scope of participatory design just to two of the four stages of standard iterative cycle of product and software development, namely to design and testing phase (excluding analysis and prototype stages) while user-centered design is limited by him only to the testing stage. Apart from some differences in definitions and scopes human-centered design is undoubtedly crucial to the development of sustainable solutions as an idea which, according to Ladner, underlies the universal design approach.

With seniors as the intended target group, it is even more important to practice co-design, since the questions of accessibility and user motivation are especially prominent. To this end Ladner \cite{ladner2015design} proposed another term: \textit{Design for Empowerment}, which involves the users in all stages of the design cycle, that is analysis, design, prototyping and testing. One difficulty with this approach when used for technological enabling solutions for the seniors is that most of them lack the ICT skills necessary to actively participate in these design stages.

\subsection{Living Labs framework concept}

Here, the concept of Living Labs comes into focus. The term \textit{Living Lab} was coined by William Mitchell from MIT \cite{niitamo2006state} and was used to refer to the real environment, like a home or a city, where routines and everyday life interactions of users and new technology can be observed and recorded to foster the process of designing new useful and acceptable products and services. Thus, the idea of LivingLab is inherently coupled with broad concept of user-centric research methodology known as human-centered design where users are no longer just subject of the functional tests but rather active participants  able to create value and contribute both content and technology. So, a LivingLab consists of an experimental environment devoted to developing and shaping new products or services in cooperation with users as co-producers or co-creators involved throughout the development process (\cite{ballon2005test,schumacher2007living}. In other words LivingLab is an open, sharing and collaborative framework rooted in the practice of participatory design which invites experimentation and facilitates the development of communities of stakeholders, enriching the product development process with surprising insights and hence, mitigating business risks.\cite{pallot2010living} These networks of partnerships ensure that Living Labs are a space where users are the focus and take part in developing solutions - therefore researching and educating the affiliated community become a viable investment.

Because the involvement of stakeholders is central to the process of co-design, the creation of technologically literate community able to, and willing to participate in the process of design is also must. To build this community Living Labs designing for seniors have to expand their focus from developing solutions, to preparing the users to participate in this process. As community training is resource-intensive it is necessary to not only develop learning solutions, but also means of allowing older adults to feel engaged and motivated, so that they stay as part of the Living Lab community of stakeholders for an extended period of time.

\subsection{Senior ICT learning}

Learning abilities are a combination of various aspects of human cognition. Unfortunately, many types of cognitive abilities are subject to age-related decline which reinforces the misconception of learning being primarily the domain of the young.

The most important factors that affect learning are an individual's information processing speed and the capacity of their working memory \cite{gamberini_cognition_????}. We have to notice that learning abilities are also influenced by a multitude of other factors which include both the learner's cognitive abilities and psychological variables such as motivation and self-confidence \cite{boulton-lewis_ageing_????}.

While the components of fluid intelligence such as working memory and acquisition of new information do indeed deteriorate with age, long-term semantic memory is generally preserved. Moreover, crystallized intelligence (associated with general knowledge, wisdom and expertise derived from accumulated experience) not only does not decline but may even benefit from aging \cite{naumanen_guiding_????}. An awareness of these age-related changes in human cognition may be the key to adapting learning programs and methods to better accommodate the needs of elderly learners. One example of such an accommodation could be the use of a learning format consisting of step-by-step instructions on performing the desired task. This has been shown to benefit senior learners, just as avoiding a lecture-based approach which has been proven to be far less effective in this age group \cite{haeggans_60s_2012}. 

\subsection{Senior motivation}

Older adults are generally considered to be more conscious of their needs than the younger generation and are interested in learning technology primarily to fulfill these needs \cite{djoub_ict_2013}. The most common uses mentioned by senior technology learners are related to utilizing ICT for communication with friends and family members (mainly via e-mail services), to supporting seniors' existing hobbies (e.g. through the use of online forums and web browsing) and dealing with everyday administrative activities such as operating a bank account \cite{boulton-lewis_ageing_????}, \cite{naumanen_practices_2008}.

Even though some seniors declare that one of their reasons for learning ICT-related skills is proving to themselves and others that they are capable of mastering technology, the majority of them are genuinely motivated by an appreciation of the benefits of using ICT in their everyday lives and consider learning it a necessity in the present times (e.g. to be able to contact and relate to their grandchildren) \cite{aula_learning_2004}.

This is why Living Lab at PJAIT first and foremost focused on developing teaching and learning activities and motivation research involving the elderly as well as the intergenerational cooperation of the, often younger, designers, application developers and seniors.

Well-designed teaching and learning activities not only enable the seniors to feel empowered and participate in the design process, but also present health benefits as staying mentally active can delay the onset of Alzheimer's and other age-related issues \cite{kotteritzsch2014adaptive}.

\section{LivingLab framework project}

LivingLab at the Polish-Japanese Academy of Information Technology is a long-term project, whose goals are to encourage social inclusion and active engagement of the elderly in social life by facilitating the development of ICT literacy among them and creating an active community of stakeholders who are both the beneficiaries and enablers of research into their problems. 
In order to expedite this and gain access to potential participants as well as a media outlet a long-term partnership with the City of Warsaw has been established, within the "Fall in love with mature Warsaw" action.

This has been initiated by offering the seniors online courses and interactive activities relevant to their needs and interests, to act as a gateway into the world of ICT solutions. Moreover, PJAIT Living Lab tests and implements crowdsourcing tools with the end goal being to encourage entrepreneurial interest and at the same time conduct research within the broad spectrum of issues connected to the problem of aging. 

To that end the e-Senior platform \footnote{$http://esenior.um.warszawa.pl/se_index.do$ site available in Polish} was created which consists of two pillars: e-learning courses and the crowdsourcing module. On this platform the registered seniors can take online courses and improve their ICT skills as well as participate in scientific research by answering surveys or signing up for offline activities. It is also the source of information about other Living Lab activities, such as location based games, Wikipedia content creation and participatory design workshops.

\subsection{On the making of LivingLab}

\subsubsection{Teaching methodology}

The Nomad e-learning system formed the foundation of the e-Senior platform. The Nomad system a comprehensive e-learning solution developed and implemented by the Polish-Japanese Academy of Information Technology.  \footnote{$https://kursy.nomad.pja.edu.pl/$ site available in Polish and English; content for Polish students as for now}

\begin{figure}
\centering
\includegraphics[height=150px]{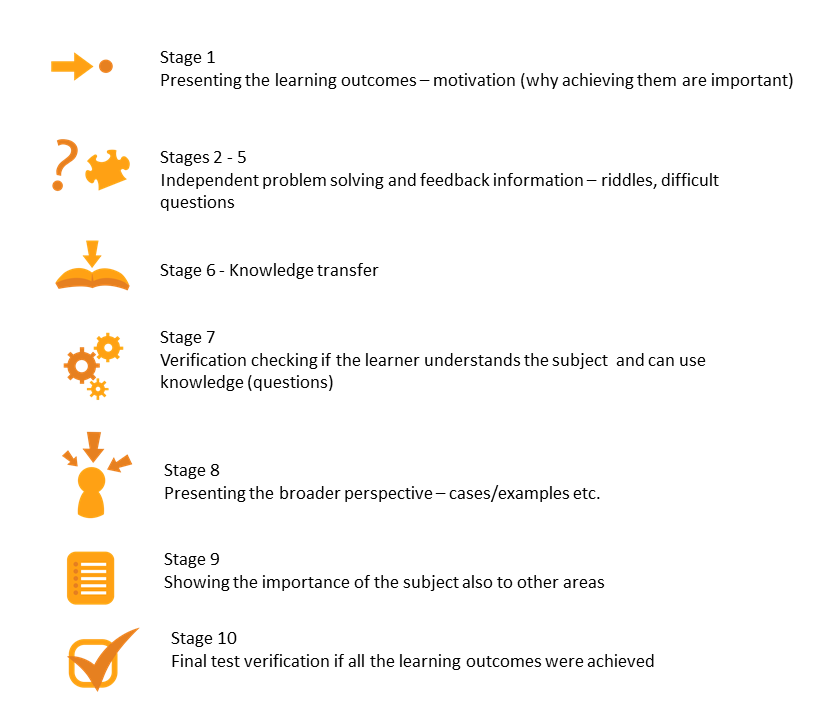}
\caption{PJAIT Nomad system -- e-learning process stages}
\label{fig:nomad}
\end{figure}

The teaching and learning methodology which underlies the design of the Nomad platform was informed by practices, such as predefined flow and learning process control. It incorporates interactive puzzles (teaser) recommended to increase engagement and make the effects of learning stick. The content creation process involves a few stages, with predefined roles for quality assurance (author, editor, reviewer and publisher) and is available with the use of an easy drag and drop interface to allow educators to create courses without knowledge of programming.

However, as often the case with online courses, the learning path appeared too rigid. The modification of adding a feature of a free-strucutred course without strict validation took many months. The learning process is oriented on progress and certification.

\subsubsection{Usability tests}

The e-Senior e-learning module and LivingLab crowdsourcing tasks module was tailored to the needs of the target group. We conducted series of usability tests with the users of the early prototype of LivingLab platform. The usability tests were performed along the predefined scheme with the use of the thinking aloud method to enhance the moderator's observation. In total six user took part in evaluating platform. 
Evaluation sessions were conducted along the same protocol and lasted about an hour. Each session consisted of a complete walktrough of typical users' activities on the platform. During the session participant's voice and face image were recorded alongside with the computer screen with the use of the Camtasia Studio software.

\subsection{E-courses module}

The e-learning courses available on the platform are designed using human-centered approach and answer older adults' needs in the area of ICT i.e. using the Internet, computer and electronic devices. They also make it easier for seniors to remain mentally stimulated by the process of learning new skills and performing interactive puzzle activities as well as solving progress tests. While the platform offers e-learning content, it lacks a usable community-building feature which, for senior citizens who are not social media users, is a drawback.

\begin{figure}
\centering
\includegraphics[height=150px]{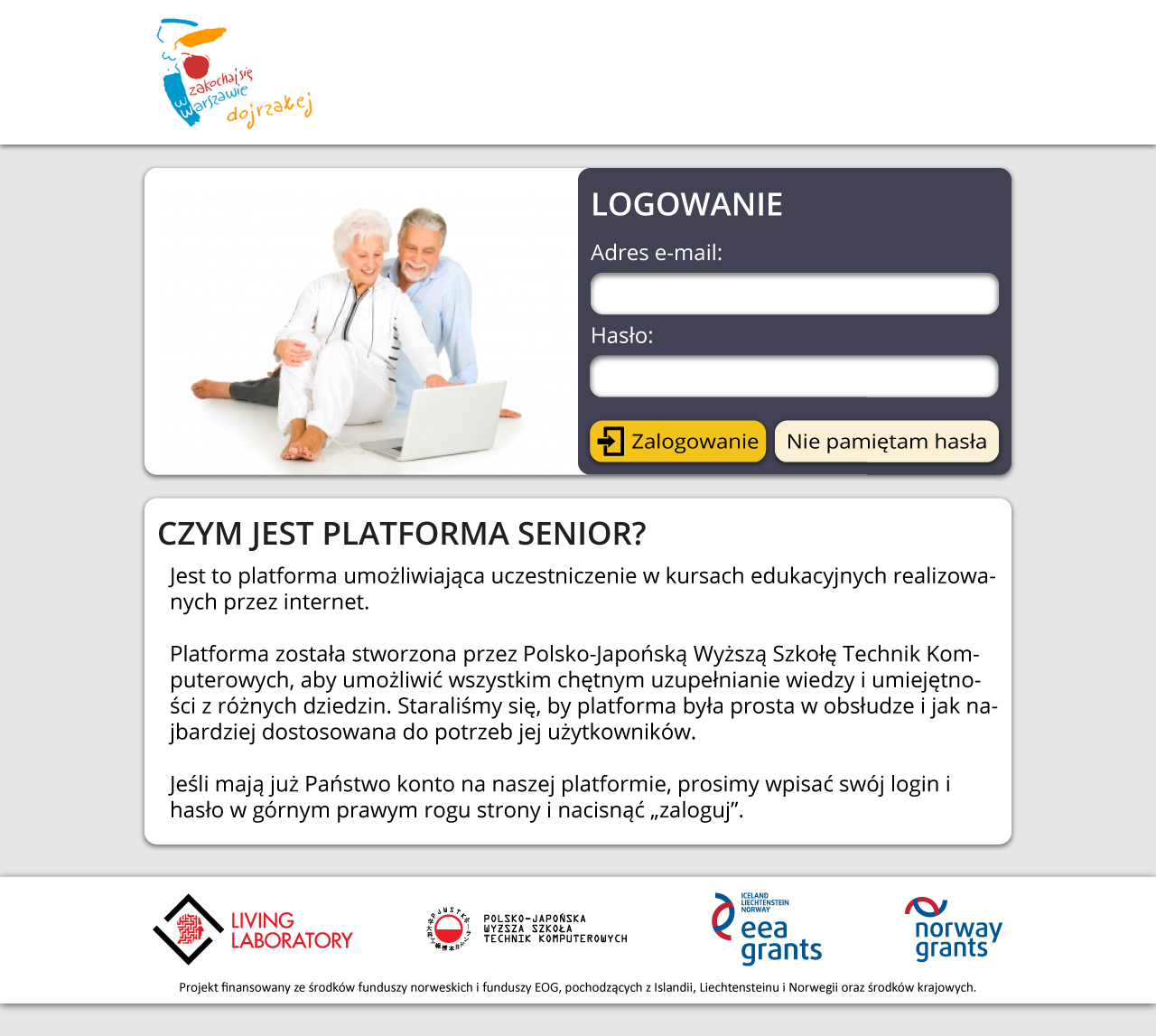}
\caption{e-Senior logon screen}
\label{fig:esenior_logon}
\end{figure}

\subsection{Crowdsourcing tasks module}

The crowdsourcing section lists research activities aimed towards the 60+ demographic in which they can take part. These can be both scientific and entrepreneurial and involve testing the platform, software designed for the elderly, filling out profiling surveys and participating in market research.

Crowdsourcing is a growing opportunity for both learning and productivity of older adults. There is a lot of proposed taxonomy of crowdsourcing, like in \cite{geiger2011managing}. There are many proposed dimensions in literature that can be related to process, task or stakeholders. In our LivingLab we refer to the approach presented in \cite{schenk2011towards}, where authors propose characteristics of three types of task: simple, complex and creative. We distinguish two kind of tasks: predefined and free flow tasks.

Predefined tasks are strictly described and can be taken or assigned to workers. The most important here is assignment, not some greater goal. There is usually strictly described criteria and end result of such task. Those are similar to simple and complex tasks from \cite{schenk2011towards}. Examples of such tasks are those in ReCaptcha, OpenStreetMap, Amazon mTurk.

Free flow tasks, where tasks can be not described, criteria and end results can be undefined and participation in such tasks is usually open for everyone and not "assigned". Here, more important is goal of collective work. This include "creative" tasks from \cite{schenk2011towards}, but in slightly different meaning described before. Those are often (but not only) community-oriented tasks \cite{rouse2010preliminary}, where participants are building some kind knowledge base, common value etc. Examples of such tasks are participating in question and answer portals (Yahoo Answers, Stack Overflow)
or various human-centered activities from user content creation like editing Wikipedia to complex and creative tasks related to software development process with the use of participatory approach.

\begin{figure}
\centering
\includegraphics[height=150px]{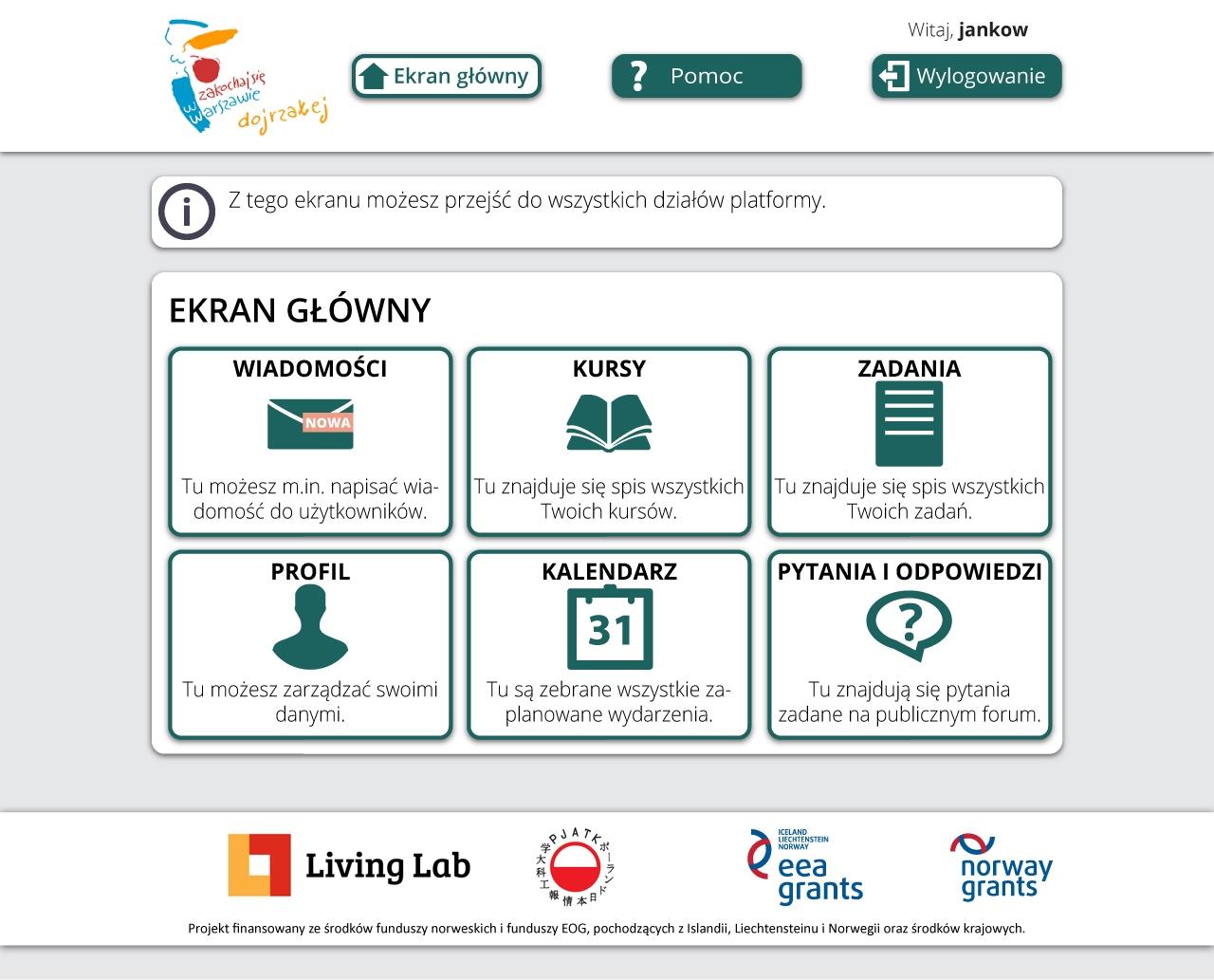}
\caption{e-Senior main panel}
\label{fig:esenior_main}
\end{figure}

\section{Activities}

The two pillars described above have been incorporated and are being tested and developed. In this section we present the data and conclusions concerning the performance of educational and crowdsourcing programs available on the platform.

\subsection{E-courses and quizzes}

The courses are based on the model developed for the Nomad platform but with adjusted controls and design.

The e-Senior platform users are recruited thanks to the long term strategic framework agreement between the city of Warsaw and PJAIT, which provides support in technologically literate community building until the year 2020. Moreover, PJAIT Living Lab cooperates with numerous NGOs. This ensures that there are more seniors willing to participate than there are available places in the courses.

The educational cycle offered to participating seniors involves stationary computer courses, which are concluded by a ICT skills test. Those who pass the test are introduced to the platform, while every participating senior can request private IT consultations. 

\begin{figure}
\centering
\includegraphics[height=140px]{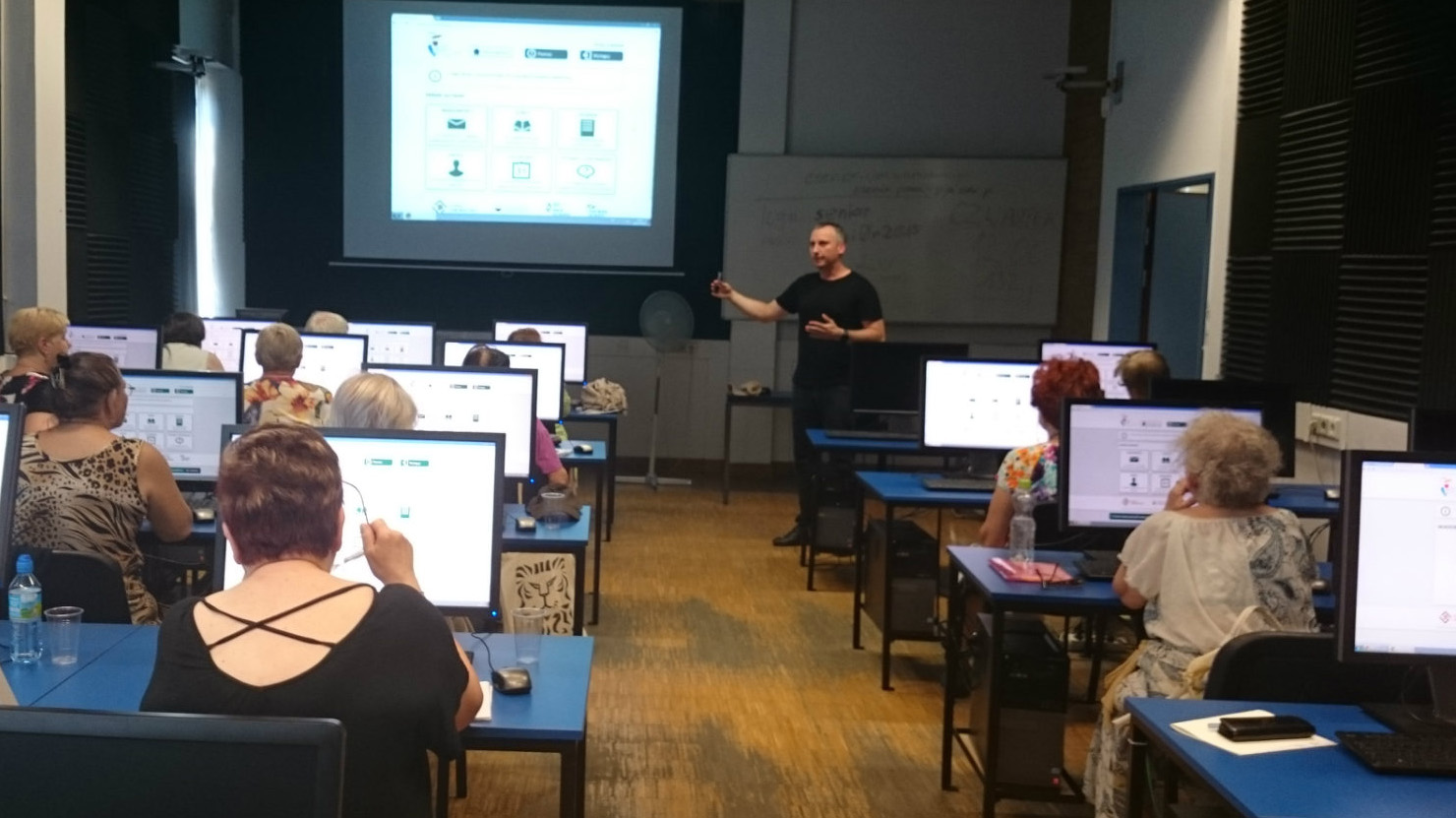}
\caption{Introductory workshops for seniors}
\label{fig:introductory}
\end{figure}

In 2016 we have started the process of modifying the offered courses to provide learning in smaller chunks and to enable non-linear education with learning pills. To enable this the learning platform had to be modified to allow for a less rigid creation of the learning path for the students based on profiling, complex event processing and incorporating a motivational system to follow the existing motivational frameworks.

\begin{figure}
\centering
\includegraphics[height=150px]{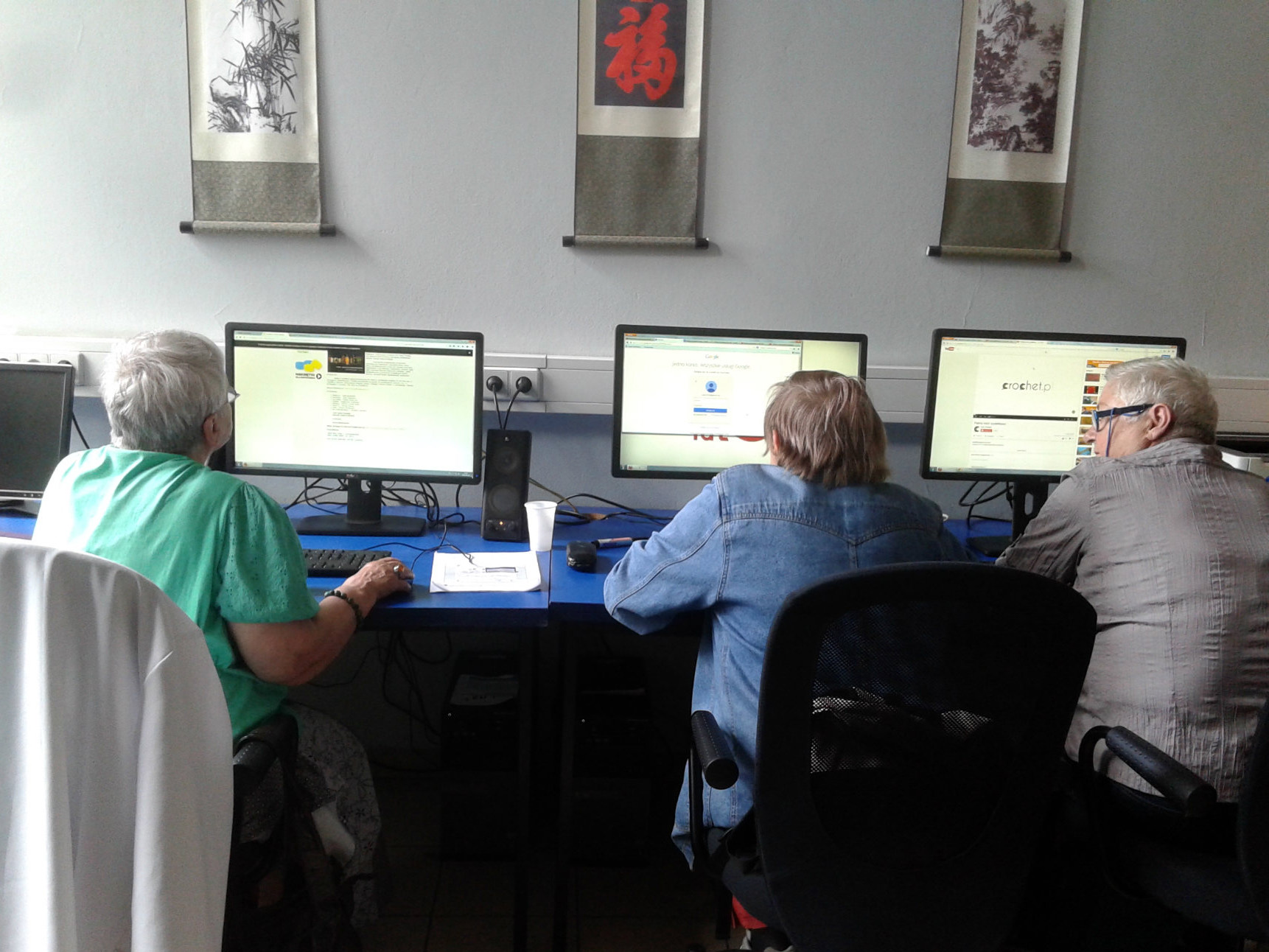}
\caption{Consultations for seniors}
\label{fig:consultations}
\end{figure}

Since 2017 we have been researching motivational frameworks for older adults, which we intend to use in the next edition of the PJAIT Living Lab. In the LL 2.0 version we intend to improve the blended learning cycle by offering longer introductory courses, complimented by consultations. The content of the courses was found too difficult and scarce. Therefore in LL 2.0 we intend to address this issue and to provide ways for users to interact via the platform, which was a common request and a key consideration to help build a community.

\subsection{Crowdsourcing}

The crowdsourcing module was to offer surveys and studies, as well as translation tasks with research, commercial and business applications.

This integrated module for web-based survey and crowdsourcing tasks has allowed over 200 participants to take profiling surveys and also to conduct an anecdotal evidence study of over 160 older participants and corresponding over 160 younger participants. The results are currently being analized and are the subject of being reported in the future.

\section{Other activities}

\subsection{Location based game}

As part of Warsaw Senior's Week Living Lab organized a location based game entitled "Stroll around yesterday" informed by best practices in game making and engagement of the elderly in city activities. The elderly  and younger tech-minded participants were divided into 15 teams of two and had to solve a mystery of a mad scientist time-shifting buildings from the communist times into the contemporary Warsaw. The game was staged to require close cooperation of team members, with the seniors using tablets to perform various ICT tasks, such as scanning QR codes, taking panoramic photos or connecting to Wi-Fi  with the assistance of their temamates given on need-to basis, and the younger participants benefiting from the background knowledge about the city and its history of the elderly. 

\begin{figure}
\centering
\includegraphics[height=150px]{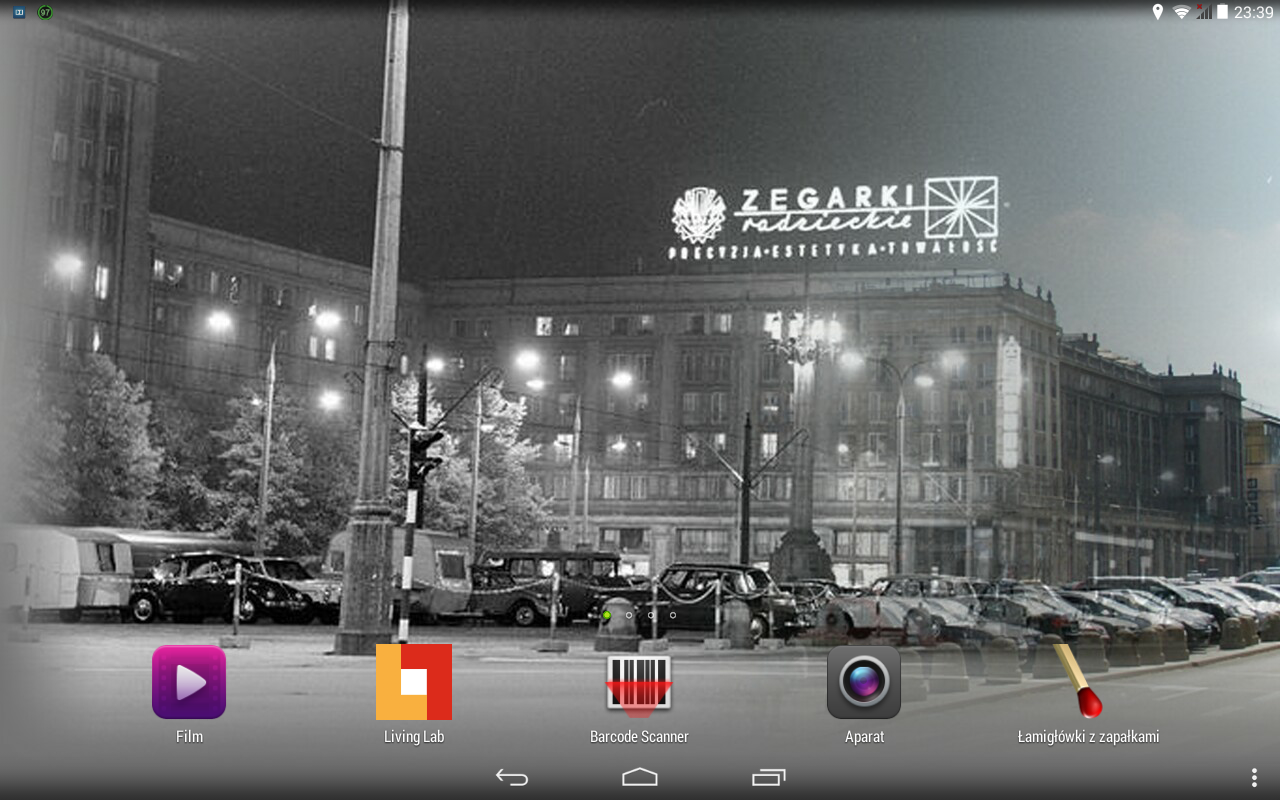}
\caption{Location-based game main screen}
\label{fig:game_main}
\end{figure}

\begin{figure}
\centering
\includegraphics[height=150px]{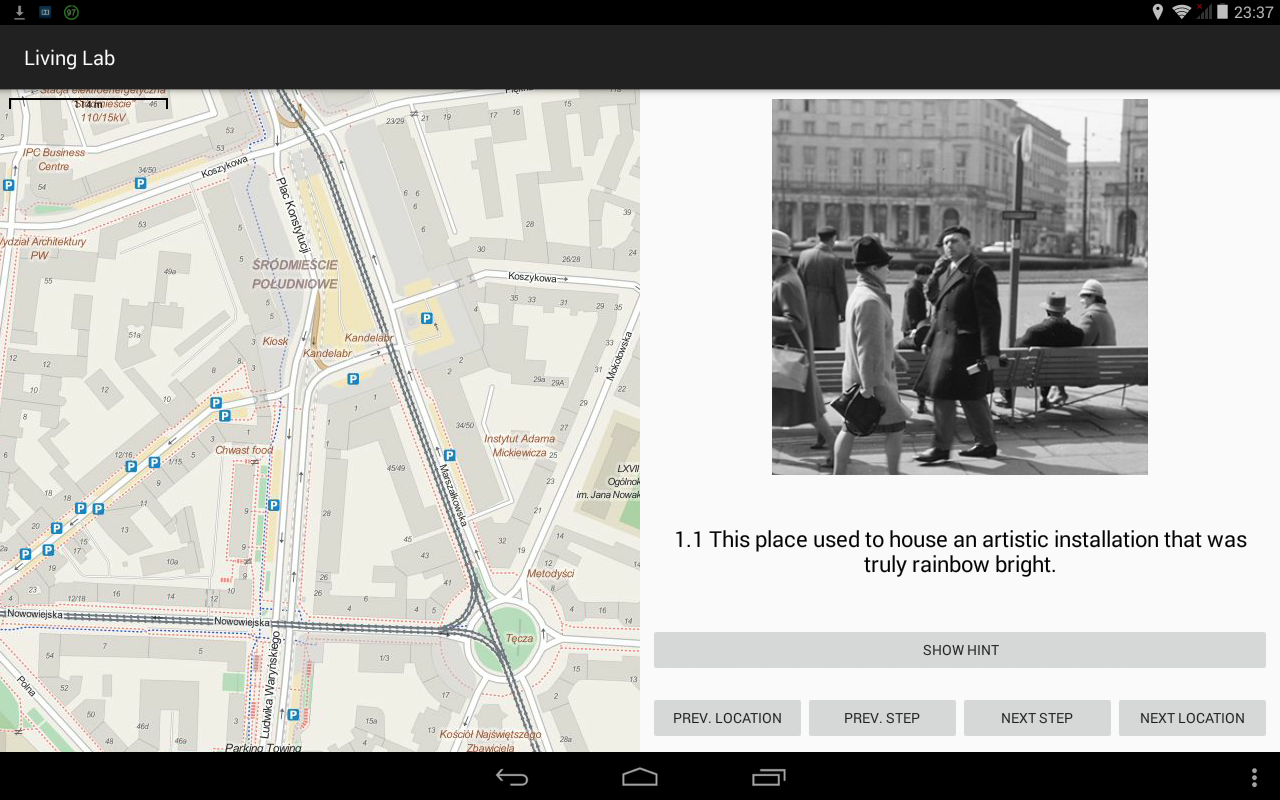}
\caption{Location-based game mobile application}
\label{fig:game_app}
\end{figure}

Overall, the game has improved the perception of the usability of mobile devices, the ICT skills of the elderly and has challenged stereotypes connected to aging.

\begin{figure}
\centering
\includegraphics[height=150px]{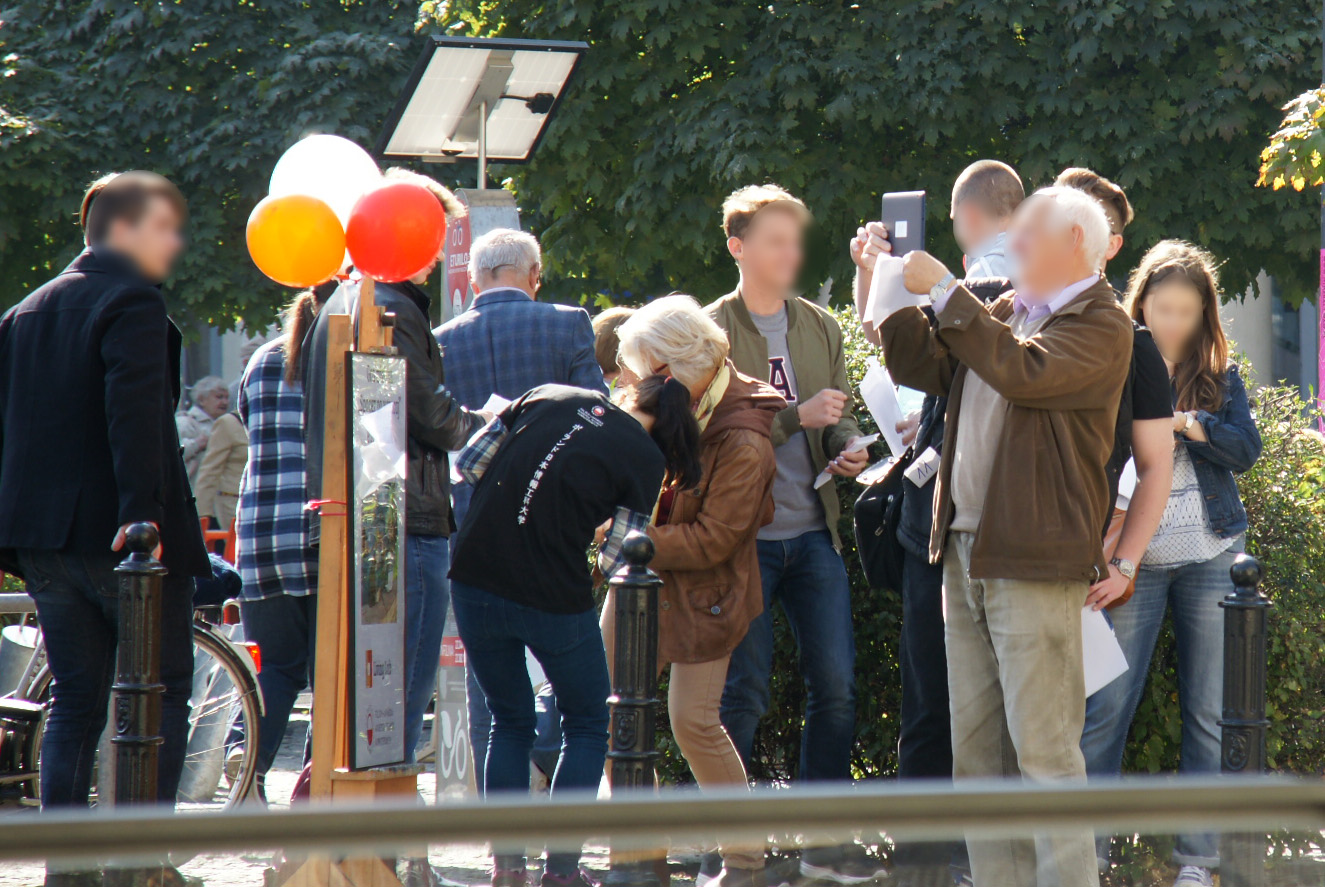}
\caption{Location-based game -- taking a panoramic photo}
\label{fig:game_panoramic}
\end{figure}

It has also provided insights into the dynamics of cooperation in intergenerational pairs and gamification as means of encouraging ICT literacy among the elderly. The results and findings from initial games was presented on GOWELL conference held in 2016 in Budapest and reported in conference precedings \cite{kopec2017location} . A quick gameplay overview is also available on-line{\footnote{$https://youtu.be/htNieG0FwfY$}}.

\subsection{Hackathons}

In order to explore the area of potential best practices in designing for and with the seniors we have organized a hackaton with older adults, which took place in March 2016. We choose this special form of case study to examine in concise form a variety of emerging strategies from user-centered design perspective to fully participatory approach. Over a hundred participants took part in the case study, including 15 older adults from our LivingLab. We used various research methods and techniques, including direct observations, surveys and in-depth interviews (both individual and group) before and after the hackathon.

\begin{figure}
\centering
\includegraphics[height=150px]{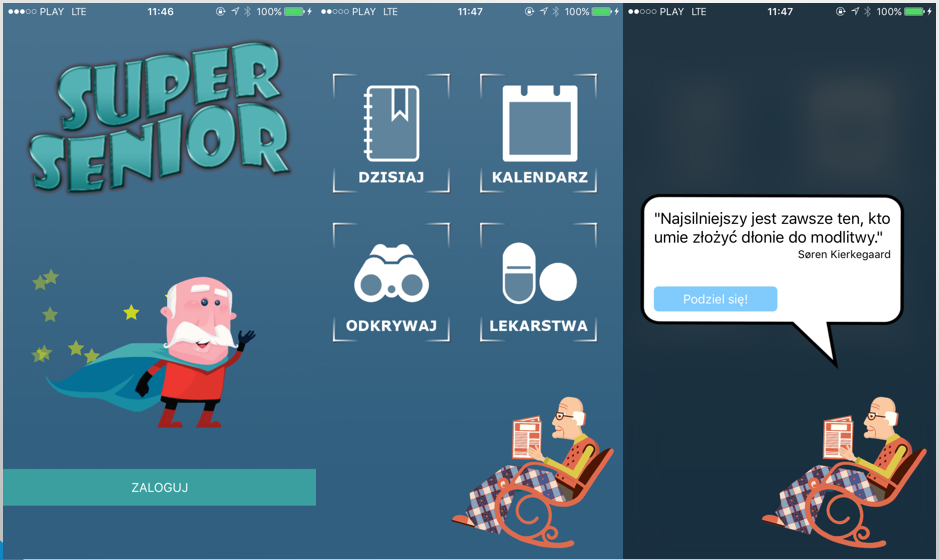}
\caption{Hackathon -- screenshot of the winning app}
\label{fig:hackathon_photo}
\end{figure}

The results of the hackathon inform the design of further studies which concern the seniors, taking into account their strengths and limitations, the best time scope and levels of engagement. 

\begin{figure}
\centering
\includegraphics[height=150px]{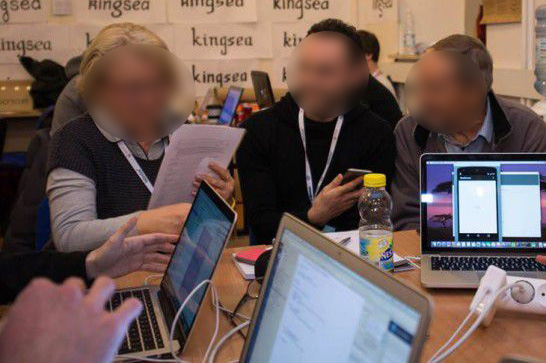}
\caption{Hackathon -- a team with seniors at work}
\label{fig:hackathon_screen}
\end{figure}

The next event is planned in 2017 with the use of findings enclosed in the extensive in-depth report from our hackathon case study submitted to Empirical Software Engineering (review in progress) and undergoing research on participatory interface design (see section below). The brief overview of the hackathon is available on-line. \footnote{$https://youtu.be/13aEBkrxe20$}

\subsection{Wikipedia workshops}

deTo further engage the seniors and to test the applicability of employing crowdsourcing solutions in this age group we organized a Wikipedia editing workshop for them. In this case the participatory process was focused on the content co-creation mechanism in connection with identifying barriers in the user interface and other limitation in better user involvement in the community driven content creation processes. 

\begin{figure}
\centering
\includegraphics[height=150px]{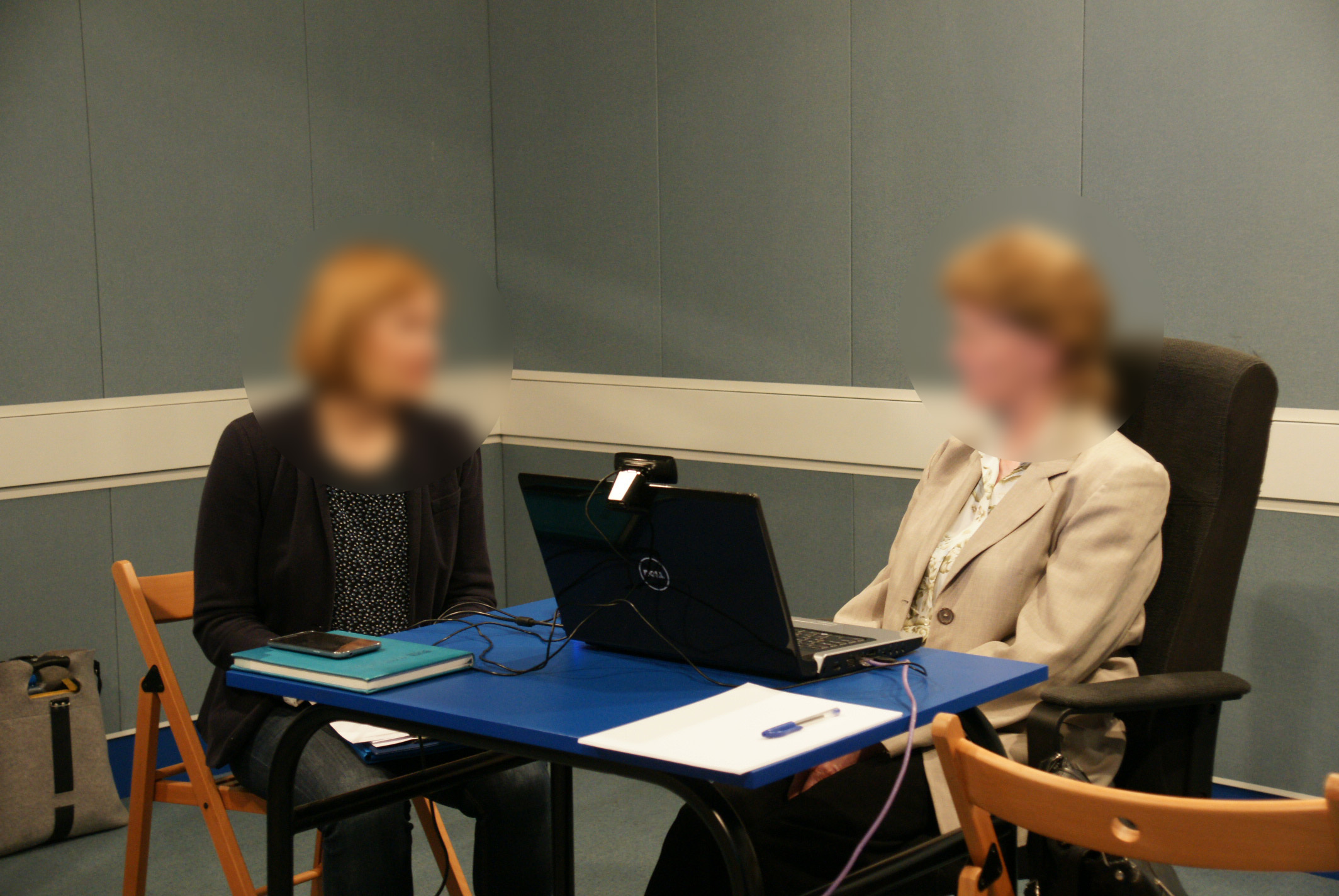}
\caption{Hackathon -- screenshot of the winning app}
\label{fig:wiki_photo}
\end{figure}

The setup of the workshop was experimental with pre-selected example tasks, which limits insight on the influence of motivation coming from creating original content and autonomous topic selection process. The workshop, however, produced insights into the senior learning process and the design of content creation tools for older adults. The detailed information of the Wikipedia study is the subject of the main WI 2017 conference call.

\subsection{User Interface Design Workshops}

In order to obtain a better and in-depth insight into the process of user involvement in participatory and co-creation processes we organized a series of workshops for our LivingLab seniors. The workshops were focused on user interface design and prototyping by the senior citizens themselves.

\begin{figure}
\centering
\includegraphics[height=130px]{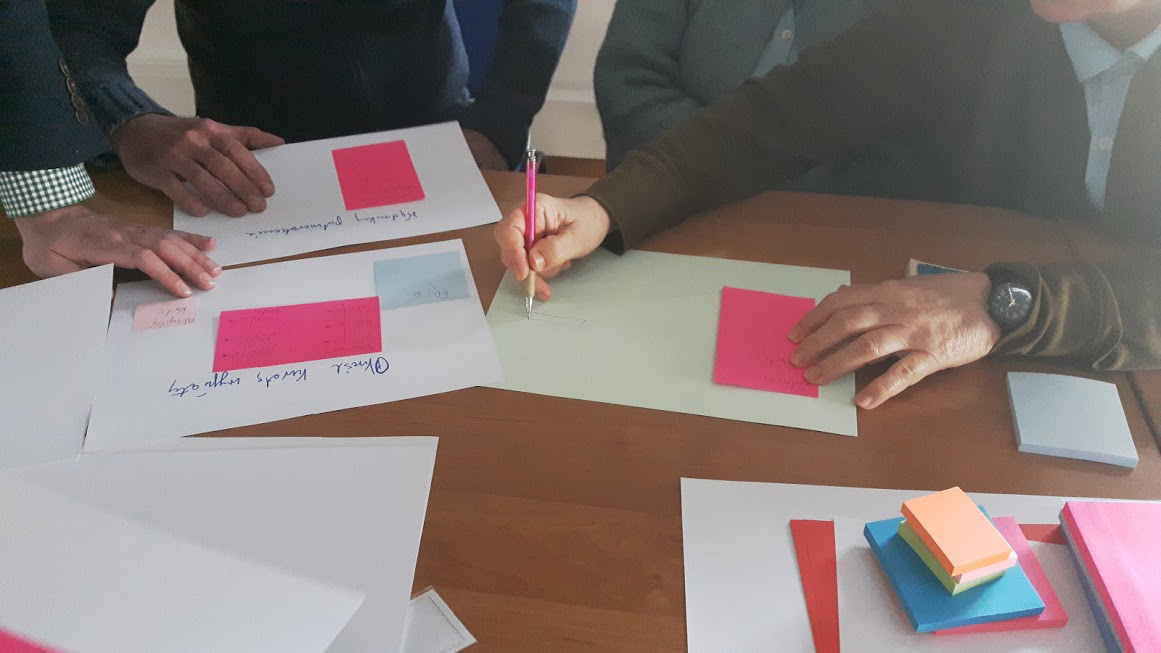}
\caption{Design workshops -- paper prototyping}
\label{fig:design_paper}
\end{figure}

Throughout the workshops we observed the older adults' prototyping learning process and identified key issues they encounter while learning abstract ideas related to human-computer interaction. We conclude the study by proposing guidelines for teaching older adults abstract technology related concepts. The outcomes of this study is the subject of another conference call. Futher in-depth research is in progress based on conclusions towards more formal guidelines for older adults effective participation in IT solutions development process.

\begin{figure}
\centering
\includegraphics[height=130px]{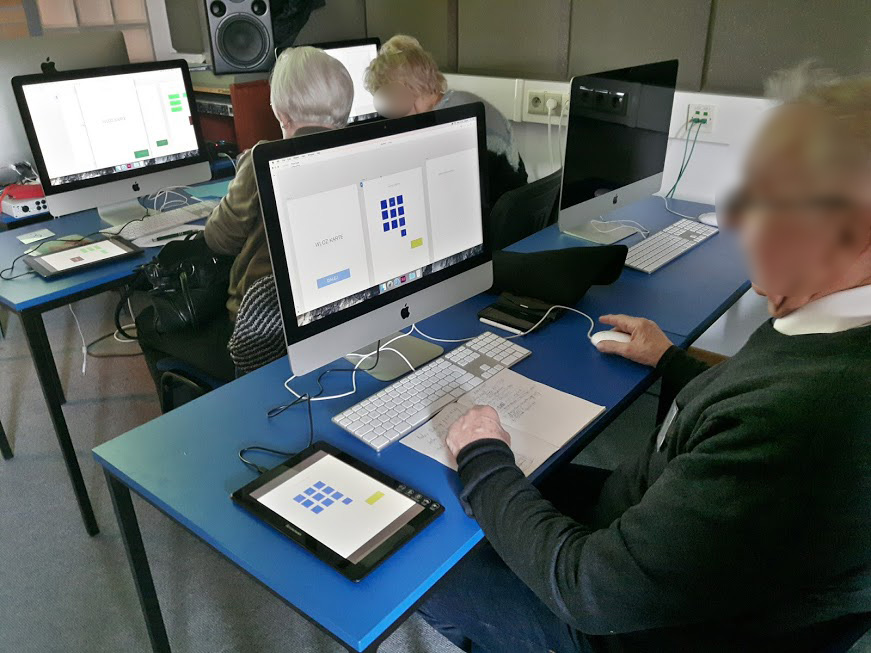}
\caption{Design workshops -- digital prototypes}
\label{fig:design_computer}
\end{figure}

\section{Discussion: lessons learned}

The activities connected to the Living Lab and the Hackatons have shown the enormous potential of the elderly as not only the users of devices and apps, but also as co-creators, who influence direction of the design process at the earliest stages from idea generation to testing prototypes. Such early involvement of the target group prevents stereotypes about their age group from influencing the end product and ensures the design and functionalities are geared towards their specific needs. 

A key insight is that the contemporary seniors prefer blended solutions, which integrate offline and online activities to standard e-learning courses and given the choice, they are more willing to participate in stationary courses and workshops. The preparatory ICT skills workshops also have to be more extensive and last for a longer time to create sustainable improvements in the way seniors interact with technology.

Regarding the crowdsourcing module, based on our observations, we concluded that one of the key motivators for them is the need to be engaged in real life social interactions and rewarding experiences, such as seeing an outcome of a game or an output produced by their team at a hackaton - otherwise the seniors get bored and lose motivation. 

Because of this it is difficult to create crowdsourcing opportunities for seniors which would make them feel sufficiently engaged. They find most of the standard crowdsourcing tasks boring and only a sense of real purpose can overcome this. According to recent Japanese research seniors have a tendency to look at a broader picture, therefore they need to clearly understand their place in the context of the task being performed in order to feel that it is worthwhile. 

Better design of social inclusion programs and applications for the elderly stemming from a better understanding of their motivation, and encouraging their participation in the whole design process would increase the rate of participation of seniors in the urban life by allowing them to be more self-assured and independent in their use of technology. To achieve this it is necessary to create a an active community fluent in ICT skills and driven by a sense of purpose.

\section{Summary and future work}

Our research and experience have shown that with prior training in ICT skills and introduction into each topic the elderly are eager to use mobile devices in order to create original content, engage in location-based augmented-reality interactions in public spaces and work in intergenerational teams. This leads us to propose that the idea of crowdsourcing solutions for urban spaces aimed at this target group has significant potential. 

We intend to further research this potential and develop best practices and content to the end of increasing senior urban participation with further strategic partnerships, one of which is in the making, involving the Federation of the Digital Universities of the Third Age - allowing us to reach seniors based outside of Warsaw, who have slightly different profiles. This will allow us to fully explore the potential of seniors as contributors to \textbf{citizen science}.

As an example of a specific future action: we have conceptualized a public space design application using geotagging and crowdsourcing. This application could incorporate photos of places which are to be redesigned, or which the application users have marked earlier as unsatisfactory. Once a photo of a public space is uploaded and geotagged the users can see it on the map. When their GPS location matches the location of the photo and they are in the presented space the possibility to re-design the space unlocks with one side of the screen showing the photo, and the other displaying drag and drop elements they can add to the photo. These could be benches, lampposts, trees, trash bins in parks, while inside public facilities they could include information signs, chairs and vending machines. Prior to using this application the seniors could participate in an e-learning course on the e-Senior platform which would teach them the basics of urban design and interior design.

To improve the appeal of the application the spaces to redesign could be connected into guided city walks with a narration, or interactive adventures where the users gather points if their designs are chosen to be implemented. This narrative dimension has proven very successful in the location-based game as the participants were motivated by the need to discover the story. 

The proposed integrated solution is promising on many levels as it may increase the amount of outdoor activity for the elderly and keep them mentally fit thanks to the need to learn and continually use ICT skills. On another level it will provide them with a way to meaningfully interact with each other and the environment. Their involvement in the pilot program will help them remove barriers to participation in technological solutions and will combat old-age stereotypes concerning ICT literacy, which may further improve their levels of happiness. Moreover thanks to the dimension of participatory design it will lead to improvements in the design of the city geared towards the needs of the elderly. 

\section{Acknowledgments}
This project has received funding from the European Union's Horizon 2020 research and innovation programme under the Marie Sklodowska-Curie grant agreement No 690962

\bibliographystyle{ACM-Reference-Format}
\bibliography{sigproc} 

\end{document}